\documentclass[12pt]{iopart}

\usepackage{iopams,graphicx}  
\usepackage[square,sort&compress]{natbib}

\begin{document}

\title[Magnetic order in single crystalline BaFe$_2$As$_2$]{Neutron scattering investigation of the magnetic order in single crystalline BaFe$_2$As$_2$}
%on magnetic order in single crystalline BaFe$_2$As$_2$}

\author{M. Kofu$^1$, Y. Qiu$^{2,3}$, Wei Bao$^4$, S.-H. Lee$^1$, S. Chang$^2$, T. Wu$^5$, G. Wu$^5$ and X. H. Chen$^5$}
\address{$^1$ Department of Physics, University of Virginia, Charlottesville, VA 22904, USA}
\address{$^2$ NIST Center for Neutron Research, National Institute of Standards 
and Technology, Gaithersburg, MD 20899, USA}
\address{$^3$ Dept.\ of Materials Science and Engineering, University of
Maryland, College Park, MD 20742, USA}
\address{$^4$ Los Alamos National Laboratory, Los Alamos, NM 87545, USA}
\address{$^5$ Hefei National Laboratory for Physical Science at Microscale and Department of Physics, University of Science and Technology of China, Hefei, Anhui 230026, China}
\ead{wbao@lanl.gov}

\begin{abstract}
%M
The magnetic structure of BaFe$_2$As$_2$ was completely determined from polycrystalline neutron diffraction measurements soon after the ThCr$_2$Si$_2$-type FeAs-based superconductors were discovered. Both the moment direction and the in-plane antiferromagnetic wavevector are along the longer $a$-axis of the orthorhombic unit cell. There is only one combined magnetostructural transition at $\sim$140 K. However, a later single-crystal neutron diffraction work reported contradicting results. Here we show neutron diffraction results from a clean single crystal sample, grown by a self-flux method, that support the original polycrystalline work.

\end{abstract}

\submitto{\NJP on Nov.\ 14, 2008. Invited contribution to the focus issue on ``Iron-Based Superconductrors''}
\maketitle

\section{Introduction}

Over the last two years, several superconductors have been discovered in fluorine-doped lanthanum oxypnictides LaFePO \cite{Kamihara2006}, LaNiPO \cite{Watanabe2007} and LaFeAsO \cite{Kamihara2008}. The common ZrCuSiAs-type (1111) structure is composed of alternating Fe(Ni)As(P) and LaO layers. Band structure calculations indicate that electronic states at the Fermi level are contributed predominately by the transition metal Fe or Ni in these semimetals \cite{bs07,A030429,A031279,A060750,A071010,A103274}, thus the importance of the layer formed by the edge-sharing Fe(Ni) pnictide tetrahedra. The same edge-sharing tetrahedral layer is also the central structural component in the previously discovered $Ln$Ni$_2$B$_2$C superconductors \cite{NiBC_str}. 
However, a distinguishing feature of the LaFeAsO system is that its superconducting transition temperature $T_C$ increases from 26 K to above 40 K \cite{A033603,A033790}, and finally reaches about 55 K at optimal F doping \cite{A042053,A042105} when La is replaced by magnetic Sm  or Ce. Other magnetic lanthanide substitutions have also been shown to result in increased $T_C$ \cite{A034234,A034283,A034384,A040835,A042582,A043727,A044290,A060926}.
Therefore, an unconventional superconducting mechanism is suspected. Indeed
while the FeP and NiP materials with their low $T_C\approx 3$-4 K may be accounted for by conventional electron-phonon interactions, the $T_C$ of the FeAs materials is too high for the phonon mechanism according to theoretical calculations \cite{A032703}. The first-principle phonon spectrum used in the calculations has since found support from neutron scattering, optical and resonant x-ray scattering measurements \cite{A051062,A051321,A073370,A073172,A073968}. 

While the $Ln$O layer ($Ln$=La, Sm, Ce, Nd, Pr, Gd, Tb or Dy \cite{Kamihara2008,A033603,A033790,A042053,A042105,A034234,A034283,A034384,A040835,A042582,A043727,A044290,A060926})
provides an excellent opportunity to investigate the interaction between superconductivity and rare-earth magnetism in the $Ln$FeAsO systems, its existence is not necessary for superconductivity. Superconductivity has been discovered in related materials with the ThCr$_2$Si$_2$-type (122) structure, where the $Ln$O layer is replaced by elemental Ba \cite{A054630}, Sr \cite{A061209,A061301} or Ca \cite{A064279} layer, and in Li$_{1-x}$FeAs \cite{A064688} and Fe$_{1+x}$(Se,Te) \cite{A072369,A074775} with the PbO-type (11) structure, which does not contain the intervening layer, but Li \cite{A072228} or excess Fe \cite{A092058} occupies an interstitial site. Therefore, the multi-orbital theoretical model based on similar semimetallic electronic states from the common Fe structural layer is likely to capture the essential physics for understanding high $T_C$ superconductivity in these Fe-based materials \cite{A032740,A033325,A033982,A034346,A041113,A044678,A061933,A063285,A101476}.  In addition, magnetic fluctuations have been proposed as the bosonic glue for Cooper pair formation. 
Currently, the extended $s$-wave superconductivity mediated by magnetic fluctuations is favored. Experiments supporting a nodeless superconducting gap have emerged \cite{A054616,A070398,A070419}.

Stoichiometric $Ln$FeAsO and $A$Fe$_2$As$_2$ ($A$ = Ba, Sr or Ca) are not superconductors. LaFeAsO experiences
a structural transition from tetragonal to orthorhombic symmetry at 150 K, which shows up as a strong anomaly in resistivity, and an antiferromagnetic transition at 137 K \cite{A040795,A043569}. For BaFe$_2$As$_2$,
the structural and magnetic transitions occur at the same temperature \cite{A062776}. The magnetic propagation vector is ($\pi,0,\pi$) in terms of the primitive tetragonal magnetic unit cell for both LaFeAsO \cite{A040795,A063878} and BaFe$_2$As$_2$ \cite{A062776}, although their crystal structures are different.
When La is replaced by magnetic Nd or Pr, the magnetic wavevector changes to ($\pi,0,0$) in the combined
Fe and rare-earth magnetic order at low temperature \cite{A062195,A074441,A074872}. For Ce substitutions, a different antiferromagnetic ordering of Ce ions was reported without the refinements provided \cite{A062528}, but the Fe part of the magnetic order is still the same as we reported for the Nd compound and is characterized by ($\pi,0,0$) \cite{A062195}.

The in-plane ($\pi,0$) magnetic wavevector is consistent with the nesting of electron and hole Fermi surfaces, which has been anticipated from band structure theory \cite{A033236,A033325,A033286}. It breaks the tetragonal symmetry of the high temperature structure and is consistent with the orthorhombic distortion at low temperature. The antiparellel moment alignment is determined by neutron diffraction to be along the longer of the in-plane axes, and the parallel alignment along the shorter axis of the orthorhombic unit cell in both NdFeAsO and BaFe$_2$As$_2$ \cite{A062195,A062776}. This magnetostriction pattern is opposite to the usual case of single orbital magnetism and is explained by careful calculations taking into account the multi-orbital origin of the antiferromagnetic order \cite{A042252}. The moment direction has also been determined to be along the longer of the in-plane axes \cite{A062195,A062776}, and the same magnetostrictive expansion and contraction have been found later in SrFe$_2$As$_2$ \cite{A070632,A071077}, CaFe$_2$As$_2$ \cite{A071525} and PrFeAsO \cite{A074441,A074872} in poly- and single-crystal studies. However, in a single crystal neutron diffraction study of BaFe$_2$As$_2$ \cite{A071743},
results different from our polycrystalline work \cite{A062776} have been reported. To clarify the issue, we show single crystal results in section~\ref{sec3}, which are consistent with our previous polycrystalline study.

\section{Experimental details}

The single crystal sample of BaFe$_2$As$_2$ was grown using a self-flux method \cite{A062452}. A distinct
feature of single crystals grown this way is that the resistivity shows a sharp drop at the phase transition at 
$\sim$140 K, similar to results from polycrystalline samples. The single crystal used by Su \etal in their single crystal neutron diffraction work was grown in Sn flux \cite{A071743}. Not only is the transition temperature much reduced, the material becomes an insulator at low temperature, in contrast to the expected metallic behavior. We conducted single crystal neutron diffraction measurements with the cold neutron triple-axis spectrometer SPINS at NIST Center for Neutron Research. The sample was mounted in a He-filled Al can in a closed cycle refrigerator so that ($h0l$) was in the scattering plane. Neutrons of 5 meV were selected using pyrolytic graphite (002) as both monochromator and analyzer. A cold Be filter was placed in the neutron path to reduce contamination from higher order neutrons. The lattice parameters are $a=5.615$, $b=5.571$ and $c=12.97\AA$ at 12.5 K in the othorhombic structure.

\section{Single crystal diffraction experiments}
\label{sec3}

\begin{figure}[t]
\includegraphics[scale=.5,angle=90,clip=true]{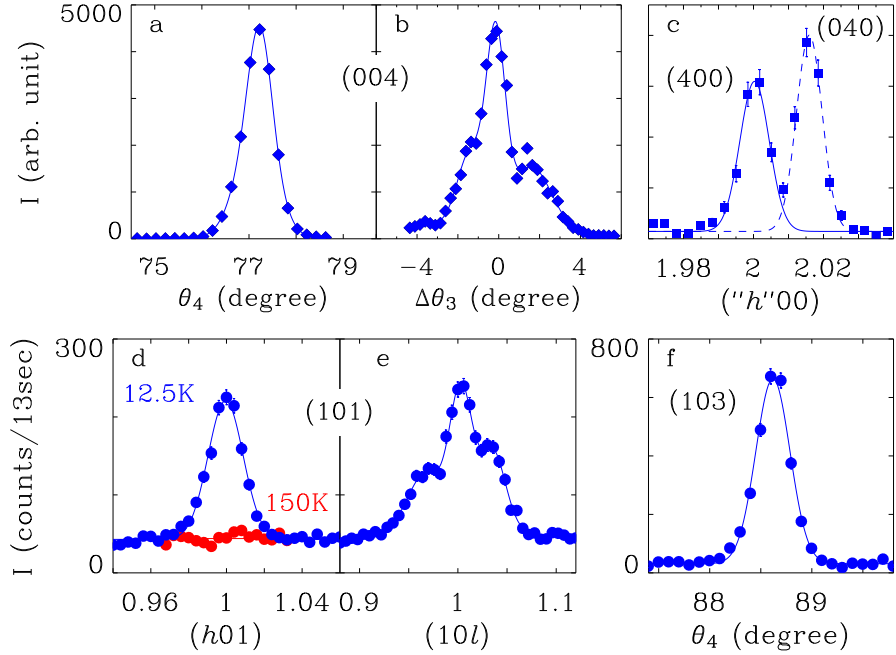}
\vskip -.2 cm
\caption{ (a-b) The $\theta-2\theta$ and rocking scans of the (004) structural Bragg peak. (c) The twinning (400) and (040) structural Bragg peaks measured using $\lambda/2$ neutrons near the (200) position. (d-e) Two perpendicular scans through the (101) magnetic Bragg peaks. (f) The (103) magnetic Bragg peak.
The blue symbols represent measurements at 12.5 K, and the red at 150 K. The error bars in all the figures represent the standard deviation in the measurement.}
\label{fig1}
\end{figure}

The mosaic of our BaFe$_2$As$_2$ single crystal sample is shown in \fref{fig1}(b). The composition uniformity is indicated by the nice peak in the $\theta-2\theta$ scan in \fref{fig1}(a). At 12.5 K, the orthorhombic distortion of the 
crystal structure is indicated by the well resolved (400) and (040) Bragg peaks due to twinning in \fref{fig1}(c). That $a>b$ is reflected in the shorter reciprocal length of the $|(400)|$ in comparison to $|(040)|$. The fact that 
only (101), not the twinning (011), peak exists in \fref{fig1}(d) is consistent with our previous determination of the (101), not (011), as the magnetic propagation vector with the definition of $a>b$ \cite{A062776}. It is opposite to what was reported by Su \etal for their Sn-flux grown single crystal sample \cite{A071743}. The $l$ scan in \fref{fig1}(e), close to the rocking direction, further supports the commensurate assignment of the  magnetic propagation vector.

Consistent with our previous data \cite{A062776}, the (103) magnetic Bragg peak in \fref{fig1}(f) is stronger than that of (101), reflecting the moment orientation factor in the magnetic neutron diffraction cross-section when the moment points along the (100) direction. At 150 K above the simultaneous magnetostructural transition,
magnetic Bragg peaks disappear completely as shown in \fref{fig1}(d) for (101).

\begin{figure}[t]
\includegraphics[scale=.6,angle=90,clip=true]{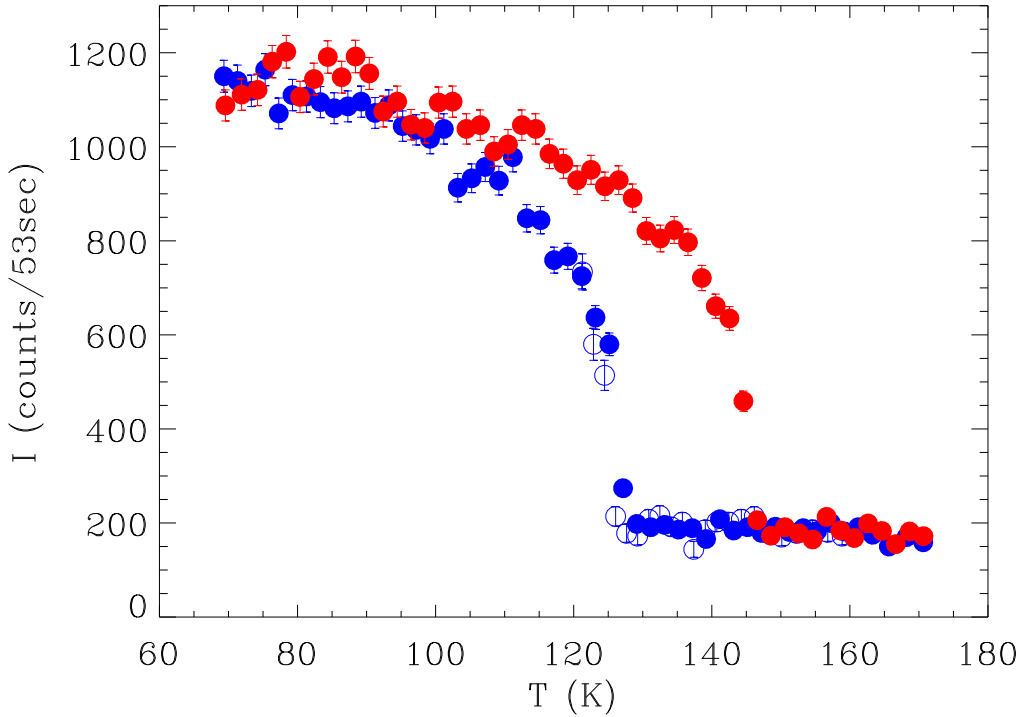}
\vskip -.2 cm
\caption{Temperature dependence of the magnetic (101) Bragg peak as a measure of the squared magnetic order parameter. The blue symbols were measured during cooling and the red heating. The temperature ramping rate was 
2.3 K/min for \fullcircle, and 4.5 K/min for \opencircle. }
\label{fig2}
\end{figure}

While there is no dispute regarding the first order nature of the structural transition in BaFe$_2$As$_2$ \cite{A062776}, the continuous appearance of the magnetic order parameter raises a scenario where the magnetic transition is of second order. If so, the lack of hysteresis in the presumed second order magnetic transition would indicate two phase transitions during a cooling/heating cycle, since obvious hysteresis in the structural transition has been observed, suggesting a similar situation to the double transition case of LaFeAsO \cite{A043569}. However, a continuous appearance of the magnetic order parameter does not preclude a magnetic transition with first-order hysteresis. This has been shown previously to occur in Ca$_3$Ru$_2$O$_7$ in a wide phase space of the temperature-magnetic field plane \cite{bao08a}, where there exists a lattice contraction associating with a Mott transition. It has also been demonstrated for CaFe$_2$As$_2$ \cite{A071525}, verifying that there is indeed only one simultaneous magnetic and structural transition in the 122 materials. In \fref{fig2}, the squared magnetic order parameter is shown during the cooling and warming cycle for the BaFe$_2$As$_2$ single crystal sample. The small difference between measurements using the two ramping rates indicates the rate is slow enough. A supercooling of about 20 K was observed, which is twice that observed previously for the polycrystalline sample \cite{A062776}. This larger hysteresis in the single crystal sample is expected due to
larger structural strain to be dissipated in the single crystal at the first order transition.

\section{Discussions}

Using a single crystal sample of BaFe$_2$As$_2$ grown by a self-flux method, the original magnetic structure determined using a polycrystal sample \cite{A062776} is confirmed.
The single crystal sample used by Su \etal has very different physical properties, likely due to inclusion of the Sn flux into the crystal. It is not clear whether the Sn inclusion is also responsible for the very different magnetic structure reported by them.
Their sample also showed decoupled magnetic and structural transitions, and the orthorhombic distortion survived at high temperature in the heating cycle. This is different from our poly and now also single crystal results. Notice that except for the results reported by Su \etal, magnetic and structural properties of the Ba, Sr and Ca 122 materials are very similar.

The prediction of the ($\pi,0$) in-plane magnetic wavevector from the nesting of quasi-two-dimensional Fermi surface before experiments \cite{A033236,A033325,A033286} certainly has boosted the credential of the spin-density-wave (SDW) mechanism for antiferromagnetism discovered in both the 1111 and 122 materials. The same SDW prediction for Fe$_{1+x}$Te \cite{A074312}, however, differs from our observed magnetic structure characterized by a completely different magnetic propagation vector ($\delta\pi,\delta\pi,\pi$) with $\delta$ tunable from 0.346 to 0.5, namely from incommensurate to commensurate magnetic structure, by excess Fe composition \cite{A092058}.
In addition, the customary energy gap from spin-density-wave order is absent in angle resolved photoemission spectroscopy (ARPES) measurements \cite{A062627}. A possible cause due to a topological constraint of degenerate orbitals has been advanced theoretically \cite{A053535}. On the other hand, there has been another theoretical approach from the strong correlation side which explains magnetic order in the Fe based materials from a localized magnetic moment picture \cite{A042252,A042480}. Both localized and itinerant theories now exist for Fe$_{1+x}$Te \cite{A094732,A103274,A111294},
and incommensurate magnetic order is also possible from either localized or itinerant perspective \cite{A042252,A104469}. Nevertheless, the strength of electronic correlations may lie in between the weak and strong correlation limits in the ferrous high $T_C$ superconductors, since electronic band structure measured by ARPES shows certain departures from the LDA band structure \cite{A062627,A070398,A070419,A072009}.

\section{Summary}

We have performed a single crystal neutron diffraction study on BaFe$_2$As$_2$. The results are completely consistent with those from our original work using a polycrystalline sample \cite{A062776}. Despite the appearance of a continuous magnetic order parameter, there is only one combined magnetostructural first-order transition in BaFe$_2$As$_2$. The conflicting results reported by Su \etal using a Sn-flux grown single crystal sample are most likely erroneous to represent intrinsic properties of BaFe$_2$As$_2$.

\ack

Work at LANL is supported by U.S.\ DOE-OS-BES, at USTC by the Natural Science Foundation of China, Ministry of Science and Technology of China (973 Project No: 2006CB601001) and by National Basic Research Program of China (2006CB922005), at UVA by the U.S. DOE through DE- 
FG02-07ER45384. The SPINS at NIST are partially supported by NSF under Agreement No. DMR-0454672.

%\section*{References}
\providecommand{\newblock}{}

%\bibliography{/home/bao/kept/tex/bib4/FeAs,/home/bao/kept/tex/bib4/mine,/home/bao/kept/tex/bib4/ruth}

\end{document}